# The Role of Wind Waves in Dynamics of the Air-Sea Interface*

## V. G. Polnikov

*A.M. Obukhov Institute for Physics of Atmosphere of the Russian Academy of Sciences Pyzhevskii lane 3, 119017 Moscow, Russia*
*e-mail: polnikov@mail.ru*
Received March 11, 2008; in final form June 19, 2008

**Abstract**—Wind waves are considered as an intermediate small-scale dynamic process at the air-sea interface, which modulates radically middle-scale dynamic processes of the boundary layers in water and air. It is shown that with the aim of a quantitative description of the impact said, one can use the numerical wind wave models which are added with the blocks of the dynamic atmosphere boundary layer (DABL) and the dynamic water upper layer (DWUL). A mathematical formalization for the problem of energy and momentum transfer from the wind to the upper ocean is given on the basis of the well known mathematical representations for mechanisms of a wind wave spectrum evolution. The problem is solved quantitatively by means of introducing special system parameters: the relative rate of the wave energy input, $I_{RE}$, and the relative rate of the wave energy dissipation, $D_{RE}$. For two simple wave-origin situations, the certain estimations for values of $I_{RE}$ and $D_{RE}$ are found, and the examples of calculating an impact of a wind sea on the characteristics of both the boundary layer of atmosphere and the water upper layer are given. The results obtained permit to state that the models of wind waves of the new (fifth) generation, which are added with the blocks of the DABL and the DWUL, could be an essential chain of the general model describing the ocean-atmosphere circulation.

**DOI:** 10.1134/S0001433809030086

## 1. INTRODUCTION

The problem of a reliable physical and mathematical describing the dynamics of the air-sea boundary (or by the other words "the air-sea interface") is the basic one, in view of solving numerous scientific and applied tasks in the problem of the middle- and large-scale interaction between atmosphere and ocean. As such tasks, they can be considered, first of all, the tasks of calculating the fields of wind, wind waves, drift currents, and describing formation of the upper water mixed layer.

In the dynamic system of the air-sea interface, the wind sea is an intermediate small-scale process. Its role in the general dynamics of the interface is rather well known at a qualitative level [1–3]. There are even existing the so called balk-formulas of different kinds, used to get relations between parameters of elements of the system atmosphere-waves-ocean [4]. But, at the level of evolution equations for these system elements, the task of describing the role of waves has not its final quantitative solution yet. This work is devoted to study the point especially.

In view of a multi-scale feature of the process of mechanical interaction between atmosphere and ocean, which is characterized by the presence of flux, wave, and turbulent motions simultaneously, this interaction is realized not directly, but by several stages. The wind waves are realized as the first pro-cess, the most clearly expressed among others, which makes later a modulating impact on both the state of the atmosphere boundary layer (ABL) and the state of the water upper layer (WUL), changing their mean parameters. Finally, the momentum and energy transfer from the wind to currents are realized by means of different direct and inverse cascade transfers in a wide band of temporal and spatial scales. Taking into account the modern knowledge about dynamics of the wind waves, one may expect that formalization of a quantitative description of dynamical processes at the air-sea interface, in terms of momentum and energy fluxes, is quite real on the basis of the wide-recognized concepts about physical mechanisms for a wind wave evolution [2, 3].

One of the main tasks of the interface dynamics is the task of calculating the mean (middle-scale) characteristics of the ABL, including the wind profile, $\mathbf{W}(\mathbf{x}, z)$, and the momentum flux to the interface, and the task of calculating the parameters of the UWL, including the drift currents, $\mathbf{U}(\mathbf{x}, z)$, and intensity of the vertical turbulent mixing. Herewith, the middle-scale wind field at the fixed horizon, for example, $\mathbf{W}_{10}(\mathbf{x})$, is considered as the given one. The closed solutions can be found from the full initial dynamic equations in particular cases only [5, 6], whilst in a general case, the detailed consideration of the system is failed [7].

The full solution of the task posed has a principal interest from scientific point of view, to get a unified understanding the dynamics of all constituents of the interface: ABL + wind waves + WUL, and to construct their qualitative models. From the practical point of view, a presence of these models opens the opportunity to solve numerous tasks related to the interface dynamics. As an example of these tasks, one can point out a self-adjusted calculation of wind, waves, and currents; calculating exchange by the heat, gas, and passive impurities between atmosphere and ocean; calculating the mixing in the upper layer, including the tasks of the impurities spreading, bubbles layer origin, and so on. It is clear that the named middle-scale processes will be modulated by one or another way due to a quickly changing small-scale wind sea. More over, the further concordance of the middle-scale processes, taking place at the interface, with the large-scale processes in atmosphere and ocean gives an opportunity to asses an impact of the wind sea state on the circulation of atmosphere and ocean as a whole.

---

* The article was translated by the authors.

A series of practically important solutions of the general task formulated above can be constructed in the frame of present mathematical concepts for mechanisms of a wind wave evolution.

Demonstrating the results of calculations testifying an impact of the wind sea state on the interface dynamics, besides the formalization of the general task, is one of the aims of present work.

## 2. MECHANISMS OF A WIND WAVE EVOLUTION AND THEIR ROLE IN THE DYNAMICS OF INTERFACE

### 2.1. General Ratios

The process of wind impact on the water upper layer is realized by means of occurrence of a turbulent layer of atmosphere at the interface, generating its wave motions. For this reason the wave motion has a stochastic feature. Therefore, description of the physical picture for wave dynamics is done by using evolution equation for a two-dimensional energy spectrum, $S \equiv S(\sigma, \theta, \mathbf{x}, t)$ [2, 3]. Here $\sigma$ and $\theta$ is the frequency and direction of propagation of a wave component with the vector $\mathbf{k}$; $\mathbf{x} = (x, y)$ is the space coordinates for wave field, $t$ is the time. In the case of deep water (neglecting the current influence on the shift of wave frequency), the equation named has the kind[1]

$$\frac{\partial S}{\partial t} + C_{gx}\frac{\partial S}{\partial x} + C_{gy}\frac{\partial S}{\partial y} = F \equiv NL + IN - DIS, \quad (1)$$

where the full derivative with time is standing in the left hand side, and the right hand side is the so called source function of the wind wave model, $F$. In the frame of approximations used, the source function $F$ includes

[1] For a brevity of the general points statement, we do not consider here the issues related to the form of evolution equation (1), which is valid for the case of presence of remarkable current.

three main terms being the constituents of the general evolution mechanism for wind waves [2, 3]:

• the rate of nonlinear energy transfer through the wave spectrum, $NL$ (the nonlinear mechanism or "nonlinearity-term";

• the rate of energy input form the wind to waves, $IN$ (the input mechanism or "input-term");

• the rate of wave energy loss, $DIS$, (the dissipation mechanisms or "dissipation-term").

Due to an extreme complexity of the system considered, an analytical representation of the source function terms can not be found theoretically even at any reasonable approximation. Therefore, in practice, it is used to involve into consideration the so called parameterizations of the terms, expressions for which are not so unequivocal. Difference of mathematical expressions of the source function terms gives the difference between existing models and defines a specificity of the physics accepted in a certain numerical model. The most known and widely spread models for wind waves of this kind are the WAM [8] and WAVE-WATCH (WW) [9] models constructed more than 10–15 years ego.

For our further calculations we shall use more modern version of the wind wave model, described in [10]. This model differs from the models said above by the source function, only, which is more physically substantiated and optimized in accordance to the criterion of fastness-accuracy of calculations. Nevertheless, despite of the named differences of the models, the general approach to the problem of description of the wind waves role in the interface dynamics is conserved unchangeable for any parameterization of the source function, because the real physics of the processes does not depend of the kind of their certain mathematical representations.

Physical sense of each term of the source function is well known and described in details, for example, in books [2, 3]. Therefore, further we shall not dwell on details of this point, restricting ourselves by specifications of the roles of wave evolution mechanisms in the process of energy redistribution among all interface elements, i.e. through the wave spectrum and between ABL and WUL.

### 2.2. The Role of the Wind Input Mechanism

The most compact approximation for the term $IN$, which is widely used till present, is written in the Mile's model representation of the form [11]

$$In = \beta(\sigma, \theta, u_*)\sigma S(\sigma, \theta). \quad (2)$$

Here $\beta(\ldots)$ is the so called wave growth increment depending on the series of the system parameters. The friction velocity, $u_*$, is the most essential among them, which is unequivocally related to the vertical turbulent flux of the horizontal momentum of the wind, $\tau$. For any horizon $z$, this relation is given by the ratios[2]:

$$u_*^2 = \tau/\rho_a = C_d(z)W^2(z), \qquad (3)$$

where $\rho_a$ is the air density, and $C_d(z)$ is the friction coefficient at the horizon $z$. Footnote that due to impossibility of detailed theoretical derivation of the explicit form for $\beta(\ldots)$, semi-empirical representations of $\beta(\sigma, \theta, u_*)$ are widely used at present.

The detailed form of $\beta(\ldots)$ is not principal for our aim. It is more important that numerous measurements show an existence of dependence for both the friction velocity and the local wind profile, $W(\mathbf{x}, z)$, on the state of wind sea, i.e. on the shape of spectrum $S \equiv S(\sigma, \theta, \mathbf{x}, t)$. Just a determination of this dependence is the task of searching for the feed back influence of wind waves on the state of ABL. It can be self-consistently solved by means of construction of a model with the so called block of dynamic boundary layer (DABL).

The said means that just the wind-waves energy exchange mechanism $IN$ is responsible for the dynamics of the ABL being considered as an element of the air-sea interface.

### 2.3. The Role of Dissipation Mechanism

This mechanism is the least studied. But the details of the dissipation function $DIS$ are not principal for our analysis of the interface dynamics. The fact is important that all partial and very quick dissipation events (breaking, instability of shear currents, sprays and foam origin, white capping, and so on) create finally a micro-scale turbulence of the water upper layer of certain intensity. The turbulence is considered as the main factor of the wave energy loss in the WUL. In a physical sense, the turbulence in the WUL is realized as an effective viscosity for motions of the wave scales. In such a case, the general formula for $DIS$ can be represented in the form

$$DIS(\sigma, \theta, S, W) = \nu_T(\sigma, \theta, S, W)k^2 S(\sigma, \theta). \qquad (4)$$

The function of an effective viscosity, $\nu_T$, is proposed to be dependent directly on the state of wave sea, for example, as a series with respect to wave spectrum $S(\sigma, \theta)$. Final result is achieved by restriction of such a series with its several first terms. The short description of this theory is given in paper [10], the final formulas are written there, as well, which are used in our further calculations related to an estimation of the dissipation processes impact on the dynamics of WUL.

Estimation of the wave energy parts, transferred into the turbulence of WUL and the drift currents, should be done with the help of a special block of dynamic upper layer (DWUL), by analogy with the block of dynamic ABL. In such a block, the tasks of description of the wind sea state, and turbulent and current motions in the WUL, are solved in a self-consistent manner, taking into account dependence of $DIS$ on the wave spectrum $S$. In a final form, such a block is in not present in modern models till now. Herewith, there are exist some simplified versions of it, based on balk-formulas relating wave parameters with parameters of the WUL (see, for example, [13, 14]).

### 2.4. The Role of the Nonlinear Evolution Mechanism for Wind Waves

This mechanism is very well studied theoretically (see bibliography in [15]). The term $NL$ of the source function $F$ is a simplified analytical representation for the Hasselmann's integral [16] which defines the nonlinear evolution mechanism. Theoretical investigations have shown that nonlinear energy transfer through the wave spectrum has a conservative feature (i.e. it does not change the total energy of waves), and it provides the wave energy transfer from a higher frequency domain of the wave spectrum into the domain located below the peak frequency of spectrum, $\sigma_p$, leading to an increase of a dominant wave length.

Finally, just the nonlinear mechanism of evolution procures a significant growth of the total wave energy in the course of their development, as far as just long waves can have a great height (and energy) without breaking and dissipation.

Due to progressive feature of waves, the energy gained is gone from the region of its income to waves into the region of its farther propagation. Therefore, at a certain point of the field, the wave energy is determined by not a local wind but by the whole wind field, where waves are generated and can come to this point. It means that the wave energy at this point is determined by the energy exchange dynamics on the whole domain of wave propagation through the space under consideration. This fact shows the principal role of term $NL$ in the wave dynamics and, consequently, in the dynamics of energy distribution between items of the interface.

## 3. INTEGRAL CHARACTERISTICS OF THE INTERFACE

### 3.1. General Ratios

The wind is the energy source for all kinds of mechanical motions realized in the items of air-sea interface. Denoting the local middle-scale wind at the standard horizon as $\mathbf{W}$, let us write expressions for the key characteristics of the wind flux. In particular, the surface density of the energy flux of local wind is[3]

$$\mathbf{F}_{WE} = \rho_a W^2 \mathbf{W}/2, \qquad (5)$$

and the local density of the flux of horizontal momentum in given by the ratio

$$\mathbf{F}_{WM} = \rho_a W \mathbf{W}. \qquad (6)$$

---

[2] Here we omit the sing in font of T, defining the direction of the momentum flux downward in vertical coordinate $z$, measured upward from the mean sea level. The point of dependence $t(z)$ is not considered here, as well.

Here $W$ is the modulus of the local wind speed. Both these fluxes correspond to the unit air volume located at the standard horizon. Just these fluxes provide a certain rate of energy and momentum input to waves, which can be expresses via the function of input term $IN$.

Function $IN$ is determined by the wind sea state, i.e. by the spectrum $S$, and by the parameters of ABL: $\mathbf{W}_{10}$ or $u_*$. In such a case, the total energy flux from the wind to waves (per unit surface), $I_E$, is given by the expression

$$I_E = \rho_w g \int\limits_0^\infty d\sigma \int\limits_0^{2\pi} d\theta In[\mathbf{W}, S(\sigma, \theta)], \qquad (7)$$

and the momentum flux, $\tau_w$, transferred from the wind to waves, is given by the ratio [3, 6][4]

$$\tau_w(z) = \rho_w g \int\limits_0^\infty d\sigma \int\limits_0^{2\pi} d\theta_\sigma^k \Phi(k, z) In[\mathbf{W}, S(\sigma, \theta)], \qquad (8)$$

where $\Phi(k, z)$ is the vertical structure function for the wave part of momentum flux, expression for which is found from the numerical model for ABL. Further, function $\Phi(k, z)$ is supposed to be known.

It is left to add that the total value of local density for the wave energy per unit surface at the fixed time moment, $E(\mathbf{x}, t)$, has the kind[5]

$$E(\mathbf{x}, t) = \rho_w g \int\limits_0^\infty d\sigma \int\limits_0^{2\pi} d\theta S(\sigma, \theta, \mathbf{x}, t). \qquad (9)$$

Hereafter $\rho_w$ is the water density, and $g$ is the acceleration due to gravity.

In the theory of ABL over a waving surface of the interface, it is used to suppose that the full flux $\tau$ consists of several summands: the viscose part, $\tau_v$, the turbulent one, $\tau_t$, and the wave induced part, $\tau_w$. But, at the height greater than the width of the viscose sublayer (of the order of $10^{-6}$ m), the viscose item of the vertical momentum flux can be neglected. Besides, the full momentum flux can be considered as a constant value along the vertical coordinate (due to quick stabilization of the air flow [5, 6]). Thus, in the cases considered, one may write

$$\tau = \tau_t + \tau_w = u_*^2/\rho_a = \text{const.} \qquad (10)$$

Now, if one writes the wave part of the momentum flux, $\tau_w$, via the spectrum in accordance with (8) and expresses the turbulent part of momentum flux, $\tau_t$ (which is an analogue of the flux over the hard wall) via a some function of the mean wind profile, $W(z)$, then the task of description of the DABL-block (construction of the DABL-block in the model (1)) becomes closed in terms of the wave spectrum.[6]

Note that a closure of relations between small and middle scales of the system, realized by the DABL, is one of a chain of the scheme of redistribution for the energy and momentum fluxes from the wind to waves, considered here. As one could see farther, analogous consideration is acceptable with respect to parameters of WUL, as well. To this end, the following set of parameters is used.

Namely, the surface density rate of energy transfer from waves into the water upper layer is determined by the expression for the dissipation term in the source function of model (1), i.e. it has the kind

$$D_E(\mathbf{x}, t) = \rho_w g \int\limits_0^\infty d\sigma \int\limits_0^{2\pi} d\theta Dis[\mathbf{W}, S(\sigma, \theta, \mathbf{x}, t)]. \qquad (11)$$

The density of energy flux (11) is corresponded by the following momentum flux from waves into the WUL

$$\tau_d = \rho_w g \int\limits_0^\infty d\sigma \int\limits_0^{2\pi} d\theta_\sigma^k Dis[\mathbf{W}, S(\sigma, \theta)]. \qquad (12)$$

Both these fluxes (jointly with the turbulent constituent of the momentum flux, $\tau_t$, considered as a rest of the full wind stress, which was not accepted by waves) regulate an origin of the WUL turbulence and drift current, $\tilde{\mathbf{U}}(\mathbf{x}, t)$. The letter defines the surface density flux of kinetic energy due to drift currents, $E_C = \rho_w \tilde{U} U^2/2$, whilst the WUL turbulence intensity is done by the rate of turbulent energy production per unit surface, $E_T$. Physical peculiarity of the situation is that the turbulent and current motions, arising in the WUL, differ by their scales. The only unarguable conditions of their distinguishing are the conservation laws. Therefore, a ratio between these values is not known in advance. To its definition, it needs a construction of the dynamic model for WUL (the DWUL-block).

In full value, the model of dynamic WUL (or the DWUL-block) needs a separate multi-stage investigation. Nevertheless, the picture of the energy and momentum redistribution, coming into WUL, becomes rather clear, taking into account the role of dissipation processes in waves at the air–sea interface. It is evident, that just the mechanism of wind wave energy dissipation does play as a whole a crucial role in the dynamics of WUL, but not the local wind does only, which is still one of parameters of this mechanism.

---

[3] For simplicity of notions, the arguments of local space and time $(x, t)$ are sometimes omitted in characteristics considered.

[4] Hereafter, during the ABL dynamics description, the commonly used definitions are given [3, 4, 6].

[5] In the practice of wind wave modeling the wave energy frequently is measured in units of wave height squared, which corresponds to the lack of multiplicand $(\rho_w g)$ in formula (9).

[6] One version of the DABL-block construction is given in [17] in details.

### 3.2. Specification of Numerical Calculations

To find certain estimations of integral fluxes $I_E$ and $D_E$, it needs to carry out calculations for wind wave evolution in accordance with solutions of equation (1) for certain terms of the source function. To this end, we use the formulas for the source function terms given in paper [10] and the wave spectrum calculation technique described there in detail.

Calculations of wave spectrum evolution were done for a typical frequency-angular grid and on the modeling site including 30 nodes through the axes OX and three lines of the axes OY. Spectrum on the central line of OY was calculated by the model (1), whilst values of the spectrum on the lateral lines at each time step were corrected in accordance with the values of $S(\sigma, \theta, x, t)$ obtained on the central line. Such an approach models the wave evolution on a site with a straight and infinite shore line along OY (see [10]). The discretization values, $X$ and $Y$, applied to a space grid of a rectangular kind ($\Delta X = \Delta Y$), were strongly varied with the aim to secure a wide range of non-dimensional fetches $\tilde{X}$, given by the ratio

$$\tilde{X} = \frac{Xg}{W^2}, \qquad (13)$$

where $W$ corresponds to the wind speed at the horizon 10 m. Values of $W$ were varied in the limits 10–30 m/s.

To the aim of model features demonstration, a series of simplest (academic) wave-origin conditions is quite acceptable. They are characterized by full a priory control of input wind parameters and wave parameters (see below). In such a case, initial conditions, as usual, were corresponding to a weak wind sea, given by the Pierson—Moscovitz spectrum with values $\sigma_p = (2-3)g/W$, and calculations were done till a stationary wave stage. Time step was varied in accordance with the space step. Numerical scheme for equation (1) solution was the implicit upstream scheme of the first order of accuracy (details of the scheme are given in [2]).

### 3.3. Numerical Values

In this paper, first of all, we are interesting in generalized estimations for variability of integral rate of wave energy dissipation, $D_E$, given by ratio (11), under various wave evolution conditions. In addition to this, the values of integral energy input rate from wind to wave, $I_E$, were calculated for the completeness of picture. Tabulating these values, obtained on the series of precursory calculations, is just the main task of this section.

Due to direct dependence of the values mentioned above on the wave spectrum intensity $S$, one should expect a strong variability of $D_E$ and $I_E$ on the parameters of wave evolution. Therefore, it needs to introduce special non-dimensional magnitudes permitting to simplify the tabulating values of $D_E$ and $I_E$. To this aim, we apply the following considerations.

In a wide range of fetches and wind values, the dimensional value of wave energy $E$ is varied in very wide limits. The range of variability for the wave spectrum intensity, $S$, is of the same strength. With the aim to compress the volume of information said, in the simplest case of direct fetches (i.e. in the cases when the wind has constant values and directions), it is accepted to use a generalized representation of wave characteristics distribution via the dependence of non-dimensional energy, $\tilde{E}$, and non-dimensional peak frequency, $\tilde{\sigma}_p$, on the non-dimensional values of fetch, $\tilde{X}$. The values of $\tilde{E}$ and $\tilde{\sigma}_p$ are given by the ratios $\tilde{E} = \dfrac{Eg^2}{W^4}$ and $\tilde{\sigma}_p = \dfrac{\sigma W}{g}$. In such a case, in a wide range of fetches and wind values, the stationary values of $\tilde{E}$ and $\tilde{\sigma}_p$ vary in a well defined interval of significances [2, 3]. If necessary, the dependence of dimensional values $E$ and $\sigma_p$ on the values of wind, $W$, and fetch, $X$, are easily defined, in this case, by means of simple transition from non-dimensional values for $\tilde{E}$ and $\tilde{\sigma}_p$ to the dimensional ones, $E$ and $\sigma_p$, using the ratios given above.

Now, following to the aim posed, we introduce the following non-dimensional magnitudes

$$D_{RE} = D_E/Ef_p \qquad (14)$$

and

$$I_{RE} = I_E/Ef_p, \qquad (15)$$

where $E$ and $f_p = \sigma_p/2\pi$ is the dimensional wave energy and the cyclic peak frequency, respectively. For any time moment and place of location, the non-dimensional magnitudes mentioned above gives the values of correspondent losses and inputs of energy for the period of main wave, normalized by the local wave energy. For this reason, the magnitude $D_{RE}$ may be called the relative dissipation rate (RDR), and the magnitude $I_{RE}$ does the relative input rate (RIR).

As it will be shown farther, just magnitudes $D_{RE}$ and $I_{RE}$ have a rather small rage of variability, what is convenient for its tabulating. Transition to the dimensional values $D_E$ and $I_R$, which are necessary for solution of the after following tasks, is easy carried out for any wave conditions, as far as the local values for wave energy, $E$, and peak frequency, $f_p$, are always known from numerical calculations.

Further it is interesting to get the tabulated values of $D_{RE}$ and $I_{RE}$ as the functions of fetch and wave velocity for a series of simplest conditions of wave evolution. Such calculations permit to estimate the typical ranges of variability for RDR and RIR in real waves. Two such cases are considered below.

### 3.4. Direct Fetch

Task formulation. Spatially homogeneous and constant in time wind, $\mathbf{W}_{10} = \text{const}$, is directed normally to the shore line (line $X = 0$). Initial and boundary conditions are given according to the text in sub-section 3.2. The aim of calculations is to estimate the range of variability for magnitudes $D_{RE}$ and $I_{RE}$ at the stage of developed wind sea and to determine their dependence on fetch and wind speed.

Dependencies $D_{RE}(\tilde{X})$ and $I_{RE}(\tilde{X})$ for two values of wind speed are presented in Fig. 1. From this figure it is directly seen that the in the large range of nondimensional fetches: $10^2 \leq \tilde{X} \leq 3 \times 10^4$, the dynamic range of variability of $D_{RE}$ and $I_{RE}$ is hardly more than 3 units. But for the fetches $\tilde{X} > 10^4$, corresponding to the fully developed wind sea, these magnitudes become nearly constant and have the values

$$D_{RE}(\tilde{X}) \approx I_{RE}(\tilde{X}) \approx 0.001. \qquad (16)$$

Moreover, these limiting values do not practically depend on the wind speed. It means that they are the universal ones, and the tabulating of them for a wide range of fetches and wind speeds is rather simple, what is quite important for solution of numerous applied tasks (see Sec. 4)[7].

Note that the result (16), expressed in such a certain kind, is found in numerical experiments for the first time, and, for this reason, it needs its interpretation. The treatment of result (16) consists in the following.

First of all, with no lost of generality for consideration, we may suppose that one-dimensional spectrum for the fully developed sea can be parameterized by the well known formula [18][8]

$$S_f(\sigma) = \begin{cases} 0.01 g^2 \sigma^{-5} & \text{at} \quad \sigma \geq \sigma_p, \\ 0 & \text{at} \quad \sigma < \sigma_p, \end{cases} \qquad (17)$$

---

[7] In this connection, it is worthwhile to note the fact that, in a fetch range $10^2 < \tilde{X} < 10^4$, the variability of nondimensional energy $\tilde{E}$ is more than two orders [2, 3]

[8] Formula (17) corresponds to the so called Phillips' saturation spectrum [18].

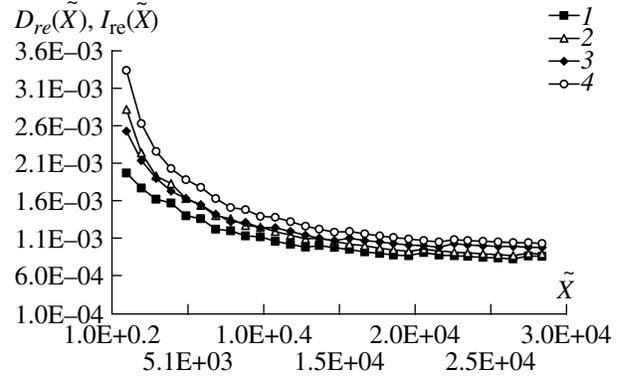

**Fig. 1.** Dependences $D_{RE}(\tilde{X})$ (lines *1* and *3*) and $I_{RE}(\tilde{X})$ (lines *2* and *4*), found for the direct fetch task. Lines *1* and *2* correspond to wind $W = 10$ m/s; lines *3* and *4* do to wind $W = 30$ m/s.

in which expression for the peak frequency is given by the ratio $\sigma_p \approx g/W_{10}$. Then, taking into account the well known empirical formula for the input term $IN$, valid in the energy containing domain, $\sigma_p \leq \sigma < 3\sigma_p$,

$$IN \approx 0.3 \frac{\rho_a}{\rho_w} \left( \frac{W_5 \sigma}{g} - 1 \right) \sigma S(\sigma), \qquad (18)$$

by means of direct integration, one may easily find

$$I_{RE} \approx \frac{\rho_a}{\rho_w} \approx 0.001, \qquad (19)$$

It means that the analytical estimation (19) of the value of $I_{RE}$ does practically coincide with the numerical one and does not depend on the wind speed, $W$. As far as for the fully developed sea the integral balance between input and dissipation should be zero, then it should be valid the balance $I_{RE} \approx D_{RE}$. Thus, the estimation presented above gives a quite convincing treatment of the result (16).

### 3.5. Swell Damping

Task formulation. The wind is absent, $W = 0$. The uniform initial wave field and the initial conditions are given in accordance with the text of item 3.2. At the boundary $X = 0$ the spectrum meaning is supposed to be constant, what corresponds to the condition for generation of a certain shape of swell spectrum at this boundary. Initial meaning of the peak frequency of swell, $f_0 \equiv f_p(0)$, is chosen in the interval of frequencies of 0.08–0.32 Hz (what corresponds approximately to the spectrum of fully developed waves for the wind speed range of 20–5 m/s). In the cases considered below, two variants are used with the following meanings of peak frequency $f_0$: (a) $f_0 = 0.09$ Hz and (b) $f_0 = 0.24$ Hz. The aim of calculation is to establish the feature of dependences for non-dimensional magnitudes $D_{RE}(X)$ and $I_{RE}(X)$ as the functions of swell propagation distance $X$, measured from the boundary of its generation. Parallel to this, the dependence is analyzed for the relative swell energy damping on the distance, $X$, given by the ratio

$$R(X) = E(X)/E(0). \tag{20}$$

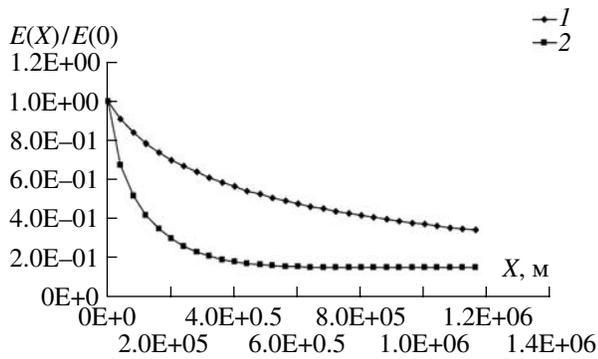

**Fig. 2.** Dependence of the relative energy of swell on the distance of propagation. (*1*) $f_0 = 0.09$ Hz, (*2*) $f_0 = 0.24$ Hz.

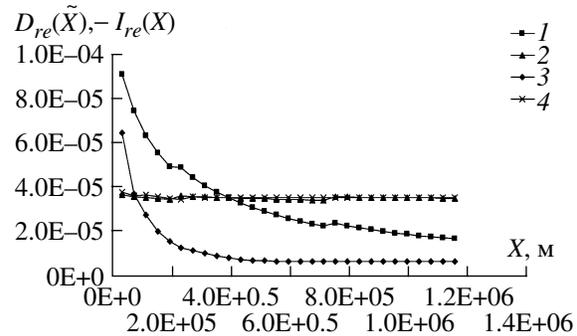

**Fig. 3.** Variability for the values of relative rates $D_{RE}$ (lines *1* and *3*) and $(-I_{RE})$ (lines *2* and *4*) with the distance of swell propagation. Lines *1* and *2* correspond to $f_0 = 0.09$ Hz, lines *3* and *4* do to $f_0 = 0.24$ Hz.

Results of calculations are presented in Figs. 2 and 3.

In Fig. 2 a behavior of $R(X)$ is shown for two different initial peak frequency values, $f_0$. It is evident an essential dependence for the swell damping rate on the value of $f_0$. Physical analysis of such a kind dependence is given in papers [3, 10], and in the present paper it is not discussed. Here it is more important to show those levels of energy damping, which are reached during propagation of swell considered, and to show the proper space scales. On this background, it is more easily to interpret the results for dependences $D_{RE}(X)$ and $I_{RE}(X)$, which are presented in Fig. 3.

In Fig. 3 it is seen, first, that the meanings of relative dissipation and input rates, in the case of swell propagation, are nearly two orders smaller than in the case of fully developed sea. On the scales of hundreds kilometers for swell propagation, numerical estimation of these magnitudes have the following meanings

$$D_{RE} \cong (1-2) \times 10^{-5}; \quad I_{RE} \cong -4 \times 10^{-5}. \tag{21}$$

Herewith, it is essential that the meaning of the rate of relative energy input, $I_{RE}$, in the case of swell propagation, is negative and independent of the value for a peak frequency of the swell. Such a result is totally secured by the kind of parameterization for the term $IN$, used in the source function of our model [10], and quite corresponds to observations (for example, [19]). Physical substantiation of this effect consists in the fact that the waves, overtaking the wind, must to return their energy into the ABL. Herewith, is should be noted that the structure of ABL must be significantly changed with respect to one for the case of swell lack, due to changing the sing for the fluxes $I_E$ and $\tau_u$. Special papers [19, 20] are devoted to discussion of such cases, where rather interesting details of the phenomenon are found, but they are not a subject of our paper.

Second, it is important to note that on the scales interesting for a practice (i.e. hundreds kilometers), the meaning of $I_{RE}$ exceeds the meaning $D_{RE}$ by the absolute value. It means that swell waves return more energy into atmosphere than do it into ocean. Therefore, despite of small values of $I_{RE}$, with the account of great distances of swell propagation, the effect of the return of energy from waves into atmosphere can leads to a radical spatial reconstruction of the large-scale air fluxes in zones, where an intensity of these fluxed were initially small. This effect becomes available to experimental registration last years only [19]. Its quantitative study needs an observation execution on a large space scales, which seems to be real not so soon.

Here it is important to note ones more that the mechanism of the inverse impact of swell waves on the ABL differs from the one responsible to impact of wind on waves, described, for example, by the Makin–Kudryavtzev model [6]. Consideration of a version of such a mechanism is given in the recent paper by the same authors [20]. In this case, it is a mechanism of energy transfer from a moving randomly roughed surface into a still atmospheric boundary layer (but not from the wind to waves). In these sense, the transfer said is close to the mechanism of wave impact on the WUL. As it was mentioned already, at present the detailed description of energy and momentum redistribution from waving surface into WUL and ABL is absent, what makes a derivation of the model for such a kind mechanism to be an actual task.

## 4. EXAMPLES FOR ESTIMATION OF WAVE IMPACT ON PARAMETERS OF THE ABL AND WUL

Thus, its was shown above that the wind wave impact on the state of ABL is realized via the energy exchange processes between wave and wind, whilst the dissipation processes in waves lead to the origin of turbulent and current motions in the water upper layer. Herewith, each of these mechanisms of wave evolution determines radically an intensity of a series of other important processes at the interface. In particular, the interface turbulence in ABL and WUL is responsible to the rate of heat and gas exchange between air and water, to the rate of passive impurity mixing and diffusion in the WUL, and so on. Besides, the wave crests breaking results in an origin of the air bubble layer in the WUL, generating in the WUL some optical and acoustic effects.

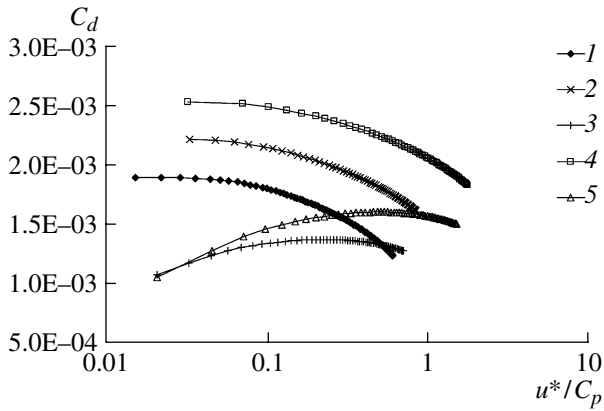

**Fig. 4.** Numerical dependencies for $C_d$ on wave age $A$ due to DBL-model of paper [3] for the series of values for wind, $W$, and spectrum shape parameter, $n$: (1) $W = 5$ m/s, $n = 4$; (2) $W = 10$ m/s, $n = 4$; (3) $W = 10$ m/s, $n = 5$; (4) $W = 20$ m/s, $n = 4$; (5) $W = 20$ m/s, $n = 5$.

Study of each the processes mentioned above needs a separate consideration. Therefore, taking as a basis our previous investigations, here we present only two examples for quantitative estimations of wave state impact on the parameters of ABL and WUL. One of them, considered in [17], deals with the calculation of the friction coefficient, $C_d$, as a function of the wind, the inverse wave age

$$A = u_* \sigma_p / g \qquad (22)$$

and the parameter $n$, describing the fall law of the two-dimensional spectrum

$$S(\sigma, \theta) \propto \sigma^{-n} \text{ for } \sigma > 3\sigma_p. \qquad (23)$$

Not all of the results, found in this direction, are known to readers, what permits to represent the most interesting of them in the text of present paper.

The second example deals with the calculation for dependence of the acoustic noise intensity due to air bubbles in the WUL on the local wind speed, $W$. It is considered in details in the recent but not well available paper [21]. Therefore, the most interesting result of this paper is also expedient here.

Additionally note that an example of wind sea impact on the drift current in a shallow water basin one may find in papers [14, 22].

### 4.1. Wind Sea Impact on the Value of a Friction Coefficient in the ABL

This issue was studied in details in paper [17]. In particular, it was noted there that an experimental variability of values for friction coefficient measured at the horizon $z = 10$ m has a dynamical range in the limits of $(0.5–2.5) \times 10^{-3}$ units, even for the fixed values of local wind, $W$. Herewith, in the case of swell, the meaning of $C_d$ can get the negative values. The letter property of the magnitude $C_d$, as it is clear at present, is totally secured by the inverse energy transfer from waves into the ABL. For this reason, below we will not dwell on this point. For the better understanding the physics of atmosphere and ocean interaction, the following question has a greater interest: what is the reason of strong variability for values of $C_d$, observable for the same wind speed, $W$? Is this effect a result of measurements errors or it is secured physically?

The answer to this question was found in paper [17], where the calculation of values of $C_d$ were executed by the use of formula (6), attracting the friction velocity values $u_*$, obtained with the DABL-model proposed in paper [6]. Analysis the results, partially represented in Fig. 4, permits to make the following conclusions: a) variability of the spectral shape $S(\sigma, \theta)$ in a high frequency domain; b) the dependence $C_d(W, A)$ is determined by the wave spectrum dynamics in the course of their developing; c) simple regression formulas for dependence of $C_d(W, A)$ does not reflect the physics of phenomenon. Such a dependence must be determined by the DABL model.

From the said it follows that a joint calculation of the ABL and the wind sea states allows not follow variability of the ABL only, but to get new physical information. Especially it is worthwhile to note that the main parameters of a wind profile in the ABL ($u_*$, $C_d$, and $W(z)$) are calculated with the DABL model directly, i.e. without attracting the hypotheses of logarithmic wind profile. On this basis, one may state that just the presence of the DABL-block leads to appearance of a new quality of the numerical wind wave model. This new quality permits to solve the applied tasks for wave forecast and atmosphere circulation with more accuracy and completeness.

To the completeness of consideration, we say several words about redistribution of turbulent and wave components of the vertical momentum flux in the ABL, found recently in paper [23]. Numerical estimations of the profile $\tau_w(z)$, made with the DABL-model said above, show that at the mean air-sea boundary, the ratio of components $\tau_t(0)$, and $\tau_w(0)$ of the total momentum flux, $\tau$, has a meaning of the order

$$\frac{\tau_t(0)}{\tau_w(0)} \cong 0.4\text{--}0.6. \qquad (24)$$

Here the value of argument $z = 0$ means a mean level of the waving surface.

Such a kind redistribution for the components of $\tau$ leads to a disturbance of the standard logarithmic profile for wind speed, depending on the wind sea state. By other words, the use of the standard logarithmic profile for wind speed and estimation of the roughness height $z_0$ with a logarithmic layer formula is a rather crude approximation to the real situation. Going out of this approximation demands a more adequate presentation of the wind profile $W(z)$, based on the use of DABL-block in a wind wave model. Consideration of this issue in more details needs a separate paper.

### 4.2. Estimation of Dependence for the Acoustic Noise Intensity in WUL on the Wind Speed

One of important practical task is an estimation of acoustic noise level, produced by the air bubbles origin in the WUL due to wave crests breaking. In particular, it is very desirable to know the dependence of bubble noise intensity on the wind speed. In a series of simplest cases, the estimation (16) for values of $D_{RE}$, obtained above, permits to give a theoretical solution of the question posed.

To this end, there was used the physical model for acoustic noise of the bubble layer in WUL. The basic ratio in this theory is the formula for intensity of bubble noise, $I_a$, of the kind [24]

$$I_a = C_T R(W, H_S) \cdot (f_r/f_0)^{-2}, \qquad (25)$$

where $C_T$ is the theoretical coefficient, $R(W, H_S)$ is the radius of bubble cloud as a function of the local wind speed, $W$, and significant wave height, $H_S$; $(f_r/f_0)$ is the non-dimensional frequency for the bubble acoustic oscillations, depending on a structure of the cloud. For further it is significant that $I_a$ is linearly dependent on the cloud radius, $R$, the value of which contains a whole information about the wind speed determining the dependence sought.

In [24] it was shown that the radius value, $R(W, H_S)$, is linearly dependent on the rate of wave energy dissipation in accordance with the ratio

$$R(W, H_S) = c_b c_t \frac{D_E}{Bh}. \qquad (26)$$

Here $D_E$ is the rate of energy income into the WUL due to wave energy dissipation, $h$ is the depth of geometrical center of a bubble cloud, and values of $c_t$, $c_b$ and $B$ are determined from experimental observations. Supposing that the values mentioned have a weak depen-dence on wind, one can get that the value $I_a(W) \propto R(W)$. Thus it can be determined on the basis of calculation for $D_E(W)$ and on physical models describing the dependence $h(W)$.

With the account of ratios (11) and (16), one may write

$$D_E(W) = 0.001 E_w(W) f_p(W). \qquad (27)$$

In case of fully developed sea, dependences $E_w(W)$ and $f_p(W)$ are given by the well known ratios (see, for example, [2, 3]) of the form $E_w \approx 3 \times 10^{-3} W^4 g^2$ and $f_p \approx g/W2\pi$. Then, under the assumptions accepted, for the acoustic noise intensity we have

$$I_a(W) \propto W^3/h(W), \qquad (28)$$

Thus, the final result is determined by the model for a depth of the bubbles cloud center.

In paper [21] there were considered different physical models for calculation of a depth for the bubbles cloud center $h(W)$ in dependence on the wind sea. They allow to state the following.

I. In the case of weak wave sea, when the bubble clouds depth has a weak dependence on the wind, dependence $I_a(W)$ is determined by the ratio $I_a \sim W^3$.

II. In the case of rather visible waves which are far from their extreme development, when the following ratios are valid: $h(W) \propto H_S(W) \propto E^{1/2}(W)$, the sought dependence (28) has the kind $I_a \sim W$.

III. And finally, in the case of fully developed sea, $h(W) \propto R(W)$, and the more reasonable dependence is $I_a \sim W^{3/2}$.

All three types of dependences $I_a(W)$ do well correspond to generalized observation data presented in Table 1. It means that the physical assumptions, used above for constructions the models of wind sea and WUL, are fairly adequate to the real processes, and the models themselves can be widely used for solution of practical tasks. The further development of the problem consists in the topic of construction the DWUL-model based on the energy and momentum balance established above for the system containing wind, waves, and water upper layer.

## 5. FINAL REMARKS

Considerations, presented above, permit to look at the whole problem of middle-scale and large-scale circulation in atmosphere and upper ocean from the new point of view. Really, up to the present, in the frame of commonly used approximations of geophysical hydrodynamics, solution of the circulation tasks was being executed without account of the waving surface state [1, 4]. In such an approach the water surface was considered as an undisturbed one (hard cover approximation), and the momentum transfer from wind into WUL was unequivocally determined by the

Empirical estimations for dependence $I_a(W)$

|      | Wind speed, m/s | Intensity $I_a(W)$ | Wave state |
|------|-----------------|--------------------|------------|
| I    | $5 < W < 10$    | $\sim\{0.004W^3 - 0.049W^2 + 0.463W - 1.5\}$ | Gravity-capillar and developing waves |
| II   | $10 < W < 15$   |                    |            |
| III  | $W > 15$        | $\sim W^{1.5}$     | Fully developed sea |

local wind. Results of investigation the role of wind waves in the dynamics of air-sea interface, presented in this paper, show that the mentioned widely spread can be radically sophisticated by means of the account of wave state, basing on numerical models including the blocks for dynamic ABL and WUL. To this aim, we have formalized the general picture for the energy and momentum redistribution in the system of air-sea interface, basing on the analysis of the roles for wind wave evolution mechanisms of wind waves.

In the frame of this scheme there is a fairly certain clarity for the wind wave model itself and for DABL-block, and even there are certain versions of them. Naturally, some details of these models can be sophisticated during their verification, but the principal approach will not get radical changes. About the model of DWUL there is not such a clarity. In particular, there is not any estimation for the fractions of energy and momentum fluxes, going from waving surface to the drift currents and turbulence in the WUL. Here it needs strong efforts to specify a model of DWUL, permitting to close the task of fluxes redistribution and to approach to solution of the middle-scale circulation task. A series of preliminary studies in this direction have been already done [14, 22]. And this fact gives a basis to expect an appearance nearest time of new wind wave models installed with the DWUL-block.

In relation to the said, it becomes possible to formulate the following addition to the present classification for wind wave models [2, 3, 25].

As it is well recognized in the world practice, the models of the third generation are ones which calculate a full 2-D wind wave spectrum, $S(\sigma, \theta)$, and have source functions operating with no limits for the wave spectrum shape (mainly, this request touches a parameterization for the term $NL$, see [25]). The models WAM, WW [8, 9] are the most widely used representatives of the models of third generation.

The model of the next generation should have a new quality. Such a kind model can be one which is installed with a special DABL-block permitting to describe a dynamic fitting of the atmospheric boundary layer to the state of wind sea, including calculation of the wind profile without involving the hypotheses of the logarithmic friction law. The model of such a level can be named as model of the forth generation. Up-to date versions of models WAM and WW, in which the friction velocity is calculated and dynamics of ABL takes place, use, nevertheless, the hypotheses of logarithmic ABL and the few-parametric representations (balk-formulas) for dependences of the ABL's parameters on the wave state's ones. Consequently, they do not meet the request formulated for the forth generation model. The model described in paper [10] does meet the request formulated above.

Following to this logics, one may state that the model installed with the block of DWUL, permitting to describe dynamics of the water upper layer (the coefficients of turbulent mixing and drift currents, at least), without involving the few-parametric dependences of them on wind and wave age, will have an additional new quality. Therefore, such a model, installed with the DWUL- block, can be classified as model of the fifth generation.

With the account of the said, the classification of wind wave models does get its logical completeness from the theoretical point of view. It is only left to realize the whole chain of models in practice. A completeness of this chain can be considered as one of the most important theoretical and practical task in the problem of joint description for atmosphere and ocean circulation at the synoptic scales. We suppose that the model [10] can serve as a basis of the task solution.

## ACKNOELEDGMENTS

The author is thankful for critical remarks and advices to Golitzyn G.S., Lavrenov I.V., Zaslavskii M.M., Kabatchenko I.M, and to the others participants of the seminars devoted to problems of the RFBR's project, No. 05-05-08027-ofi. The work was supported by the projects Nos. 04-05-64650a, 05-05-08027-ofi, 07-05-12011-ofi_a, 08-05-13524-ofi_c.